\documentclass[twocolumn,preprintnumbers,elsart]{revtex4}
\usepackage{dcolumn}
\usepackage{bm}
\usepackage{graphicx}
\usepackage{amsmath}

\begin{document}
	\title{Vortex solitons in spin-orbit-coupled atomic-molecular condensates}
	
	\author{Chao Kong$^{1,2}$, Jinqing Li$^{1,2}$, Chunfeng Huang$^{1,2}$, Yijia Zhu$^{1,2}$, Zhonghua Li$^{1,2}$, 
		Jingwen Li$^{1,2}$, Huiwen Xiang$^{1,2}$, Yuzhen Tang$^{1,2}$, 
		Zhenwang Li$^{1,2}$, Yue Liang$^{1,2}$, 
		Haiming Deng$^{1,2}$\footnotemark[1]\email{woshidenghaiming@126.com}, 
		Gui-hua Chen$^{3}$\footnotemark[2]\email{cghphys@gmail.com},  and Zhiwei Fan$^{4}$}
	\footnotetext[1]{%
	Corresponding author: woshidenghaiming@126.com} 
\footnotetext[2]{%
	Corresponding author: cghphys@gmail.com}
\address{	\mbox{$^1$ School of Physics and Electronic-Electrical Engineering, Xiangnan University, Chenzhou 423000, China}\\
	$^2$ Microelectronics and Optoelectronics Technology Key Laboratory of Hunan Higher Education, Xiangnan University, Chenzhou 423000, China\\
	\mbox{$^{3}$ School of electrical Engineering \& Intelligentization, Dongguan University of Technology, Dongguan 523000, China}\\
	\mbox{$^{4}$ School of Engineering, Newcastle University, Newcastle upon Tyne NE1 7RU, United Kingdom}}
	
	\begin{abstract}
		In our previous study of a three-dimensional (3D) atomic-molecular Bose-Einstein condensate (BEC) with quadratic nonlinearity and spin-orbit coupling (SOC), stable solitons are realized as semi-vortex (SV) or mixed-mode (MM) states, with SOC applied solely to the atomic component. In the SV state, the molecular part forms a pure vortex, while the atomic part remains a hybrid of fundamental and vortex modes. This prevents the overall self-bound state from achieving a fully pure vortex configuration, with the total vortex component always less than $100\%$. Here, we extend the study to two dimensions (2D) and uncover a qualitatively different behavior: purely quadratic 2D systems support not only ground-state (GS) SV solitons, but also higher-order excited-state (ES) solitons that are genuine full vortices (a feature absent in the 3D counterpart). For both GS and ES solitons, stable vortex states exist only when the SOC strength $\lambda$ exceeds a minimum threshold $\lambda_{\min}$, which increases monotonically with $S_1$. Conversely, the critical detuning $\alpha_c$ decreases monotonically with $S_1$, indicating that higher-order vortices are more susceptible to destabilization. For $\lambda < \lambda_{\min}$, the system forms MM states instead of vortex solitons. Our results demonstrate that dimensional reduction to 2D enables the emergence of pure-vortex excited states, establishing a new pathway for stabilizing high-charge topological excitations in ultracold molecular gases.
		
		\emph{Keywords}: atomic-molecular Bose-Einstein condensate; Spin-orbit
		coupling; Semi-vortex solitons; Excited-state solitons
	\end{abstract}
	
	\maketitle
	
	\section{Introduction}
	
	Quadratic ($\chi^{(2)}$) nonlinearity, well known for second-harmonic generation in optics and atom-molecule conversion in  Bose-Einstein
	condensate (BEC) mixtures \cite{Fejer1992,Chen1985,Duine2004,Koehler2006}, is also capable of supporting multidimensional bright solitons by suppressing collapse \cite{Torruellas19951,Torruellas1995,Rubenchik,Stegeman,HaoHe,Wise,Etrich2000,Buryak2002,Hayata1993,Corney2001,Lutsky2015,Saut,Bagci2021,B. Oles,P. D. Drummond,T. G. Vaughan}. Among these, vortex solitons carrying orbital angular momentum are of particular interest \cite{Allen1992,Firth1997,Towers2001,Trapani2000,Hang2009,Adhikari2004,Baizakov2004,Leblond2007}, yet in homogeneous media with purely quadratic nonlinearity they suffer from azimuthal instabilities that fragment the vortex ring \cite{Firth1997,LTorner,DVPetrov}. Prior efforts to overcome this obstacle have relied on auxiliary nonlinearities, such as quadratic-cubic interplay \cite{TJAlexander,DMihalache,R. DeSalvo,Wang2026}, cubic-quintic competition \cite{EMWright,M. Quiroga,I. Berezhiani,D. Mihalache,R. L. Pego,D. Mihalache2,C. Cartes,YYWang}, or quantum-fluctuation effects in quantum droplets \cite{Barbut2016,Cabrera2018,Zhao2022,Luo2021,Yogurt2023,Nie2023,He2024,Kartashov2018,Li2018,Zhang2019} (recently extended to anisotropic dipolar droplets \cite{Li2024PRL,Li2024Front}). Nevertheless, all such stabilization mechanisms require additional repulsive terms. Achieving stable vortex solitons in purely quadratic systems with a uniform background has therefore remained an open challenge \cite{VELobanov,F. Zhao,C. Kong}.

	Recently, we have proposed the first scheme for stabilizing vortex solitons in a purely quadratic nonlinear medium in free space, realized in a three-dimensional (3D) atomic-molecular BEC \cite{C. Kong2026}. In this scheme, planar spin-orbit coupling (SOC) is applied solely to the atomic part, configuring it into a semi-vortex (SV) or mixed-mode (MM) state. However, while the molecular part exists as a pure 3D vortex, the atomic part remains a hybrid (featuring a fundamental mode in one component and a vortex in the other), so the self-bound state cannot achieve a fully pure vortex configuration, with the total vortex component being less than $100\%$. Nevertheless, this 3D study leaves two key issues unexplored. First, the quantitative role of the SOC strength \(\lambda\) in the formation and stabilization of the vortex solitons. In particular, the minimum \(\lambda\) required for stabilizing a vortex, has not been systematically revealed. Second, the existence and stability of higher-order excited-state (ES) solitons with topological charges \(S_1>1\) are completely absent. On the other hand, vortex solitons supported by purely quadratic nonlinearity in uniform media are inherently subject to azimuthal instabilities in both three and two dimensions \cite{Firth1997,LTorner,DVPetrov}, a problem that does not automatically disappear upon dimensional reduction. Notably, unlike its 3D counterpart, a 2D system offers more favorable geometric constraints for vortex stabilization. However, no previous work has systematically investigated the stability of ground-state (GS) and higher-order ES solitons in purely quadratic 2D systems and presented a complete phase diagram.
	
	In this work, we systematically address this problem in a 2D  atomic-molecular BEC system. We find that, for both GS and ES solitons, stable vortex states exist only when the SOC strength $\lambda$ exceeds a minimum threshold $\lambda_{\min}$, which increases monotonically with the topological charge $S_1$; below this threshold the system forms MM states instead. In addition, we identify a critical detuning $\alpha_c$ that bounds the stability regime from above: as the detuning increases beyond $\alpha_c$, the solitons become unstable and transition into MM states. This critical detuning $\alpha_c$ decreases monotonically with increasing $S_1$, reflecting that higher-order vortex states are more susceptible to detuning-induced destabilization. The behavior of both $\lambda_{\min}$ and $\alpha_c$ is systematically characterized through the parametric dependencies of the chemical potential, energy, and angular momentum on $\alpha$ and $\lambda$. Furthermore, we find that the 2D system exhibits a capability qualitatively different from its 3D counterpart: it can stabilize not only GS vortices, but also higher-order ES solitons with topological charges up to $S_1 = 5$, a feature that is hardly achievable in 3D purely quadratic systems. These results demonstrate that the 2D atomic-molecular BEC system is not a simple dimensional reduction of the 3D case; 
	its capacity to support higher-charge vortex states offers a new physical platform for the study of high-topological-charge solitons.
	
	\section{Model}
	
	To address this issue, we propose a 2D atomic-molecular BEC with Rashba SOC applied to the atomic components. The system is modeled by coupled Gross-Pitaevskii equations (GPEs) for the atomic ($\phi_{1,2}$) and molecular ($\psi$) wave functions, which, in scaled form, read
	\begin{eqnarray}
		i\frac{\partial \phi _{1}}{\partial t} &=&-\frac{1}{2}\nabla _{\mathrm{2D}%
		}^{2}\phi _{1}+\lambda \left( \frac{\partial }{\partial x}-i\frac{\partial }{%
			\partial y}\right) \phi _{2}-\psi \phi _{2}^{\ast },  \label{+} \\
		i\frac{\partial \phi _{2}}{\partial t} &=&-\frac{1}{2}\nabla _{\mathrm{2D}%
		}^{2}\phi _{2}-\lambda \left( \frac{\partial }{\partial x}+i\frac{\partial }{%
			\partial y}\right) \phi _{1}-\psi \phi _{1}^{\ast },  \label{-} \\
		i\frac{\partial \psi }{\partial t} &=&-\frac{1}{4}\nabla _{\mathrm{2D}%
		}^{2}\psi -\phi _{1}\phi _{2}-\alpha \psi ,  \label{++}
	\end{eqnarray}%
	where $\nabla _{\mathrm{2D}}^{2}=\frac{\partial ^{2}}{\partial x^{2}}+\frac{%
		\partial ^{2}}{\partial y^{2}}$ is the 2D Laplacian, $\lambda $ is the SOC\
	strength, $\ast $ stands for the complex conjugation, and $\alpha $ is the
	mismatch of the atomic-molecular interconversion. Starting from the full 3D dimensionful equations for the atomic and molecular fields, we reduce the system to two dimensions by freezing the transverse ($z$) harmonic-oscillator ground state with common oscillator length \(a_z=\sqrt{\hbar/(m\omega_z)}\). The characteristic scales are set by the 3D healing length: \(L_0=1/\sqrt{8\pi n_0 a_s}\), \(E_0=\hbar^2/(2mL_0^2)\), and \(t_0=\hbar/E_0\), where \(n_0\) is the peak 3D density and \(a_s\) the \(s\)-wave scattering length. Introducing dimensionless variables \(\mathbf{r}=L_0\tilde{\mathbf{r}}\), \(t=t_0\tau\), \(\phi_i=\sqrt{n_{2D}}\tilde{\phi}_i\), \(\psi=\sqrt{n_{2D}}\tilde{\psi}\) with \(n_{2D}=n_0\sqrt{\pi}a_z\), we obtain equations containing a parameter \(\tilde{\chi}\) for the atom-molecule coupling. A subsequent rescaling \(\tilde{\phi}_i=\tilde{\chi}^{-2/3}\phi_i\), \(\tilde{\psi}=\tilde{\chi}^{-1/3}\psi\) absorbs this parameter, yielding the standard form (1) where the coupling coefficients are unity. The dimensionless parameters \(\lambda\) and \(\alpha\) are related to their physical counterparts by \(\lambda = \hbar\lambda_{\mathrm{phys}}/(E_0L_0)\) and \(\alpha = \hbar\alpha_{\mathrm{phys}}/E_0\), with \(\lambda_{\mathrm{phys}}\) being a velocity and \(\alpha_{\mathrm{phys}}\) an angular frequency.
	
	The system conserves its energy (Hamiltonian),
	\begin{gather}
		E=\int \left[ \frac{1}{4}\left( 2\left\vert \nabla \phi _{1}\right\vert
		^{2}+2\left\vert \nabla \phi _{2}\right\vert ^{2}+\left\vert \nabla \psi
		\right\vert ^{2}\right) \right.  \notag \\
		+\lambda \phi _{1}^{\ast }\left( \frac{\partial }{\partial x}-i\frac{%
			\partial }{\partial y}\right) \phi _{2}+\lambda \phi _{1}\left( \frac{%
			\partial }{\partial x}+i\frac{\partial }{\partial y}\right) \phi _{2}^{\ast }
		\notag \\
		\left. -\phi _{1}^{\ast }\phi _{2}^{\ast }\psi -\phi _{1}\phi _{2}\psi
		^{\ast }-\alpha \left\vert \psi \right\vert ^{2}\right] \mathrm{d}\mathbf{r,}
		\label{E}
	\end{gather}%
	with $\ast $ standing for the complex conjugate, and total norm (scaled
	number of atoms),
	\begin{equation}
		N=\int (|\phi _{1}(\mathbf{r})|^{2}+|\phi _{2}(\mathbf{r})|^{2}+2|\psi (%
		\mathbf{r})|^{2})d\mathbf{r\equiv }N_{1}+N_{2}+N_{3}\mathbf{.}  \label{N}
	\end{equation}%
	where $N_{1,2}$ and $N_{3}$ are the norms of two atomic and single molecular
	components. 
	Below, we treat $(\alpha ,\lambda )$ as a set of the system's control
	parameters. 
	
	The SOC system conserves the angular momentum \cite{Hnizdo,Scripta}. The application of the Noether theorem \cite%
	{Thompson} to the system produces the angular momentum
	\begin{equation}
		\begin{aligned} M = \int \int \Bigg\{ & \bigg[
			\frac{i}{2}\sum_{j=1,2}\phi _{j}^{\ast} \left( y\frac{\partial }{\partial
				x}-x\frac{\partial }{\partial y}\right) \phi _{j} \\ & + i\psi ^{\ast
			}\left( y\frac{\partial }{\partial x}-x\frac{\partial }{\partial y}\right)
			\psi \bigg] + \mathrm{c.c.} \\ & + \frac{1}{2}\left( \left\vert \phi
			_{1}\right\vert ^{2}- \left\vert \phi _{2}\right\vert ^{2}\right) \Bigg\}
			dxdy, \end{aligned}  \label{Mz}
	\end{equation}%
	where $\mathrm{c.c.}$ stands for the complex conjugate.

	The stability of the
	solitons was then identified with the help of systematic real-time simulations of their perturbed evolution in the framework
	of Eqs. (1-3). Accordingly, the time dependence of the wave functions of the SV states is
	fixed as%
	\begin{eqnarray}
		\phi _{1}(r,\theta ,t)=\exp \left( -i\mu t+iS_1\theta\right) \Phi _{1}(r),  \notag \\
		\phi _{2}(r,\theta ,t)=\exp \left( -i\mu t+iS_2\theta \right) \Phi _{2}(r),
		\notag \\
		\psi (r,\theta ,t)=\exp \left( -2i\mu t+iS_3\theta \right) \Psi (r),  \label{mu}
	\end{eqnarray}%
	where the chemical potential of the atomic components is
		$\mu \leq \mu _{\max }=-\lambda ^{2}/2$ \cite{Sakaguchi}, and the integers $S_1$, $S_2$, $S_3$ obey the angular-momentum matching conditions $S_2 = S_1 + 1$ and $S_3 = S_1 + S_2$. With $\Phi _{1,2}(r)$ and $\Psi (r)$\ being real functions,
	the corresponding radial equations are given by%
	\begin{eqnarray}
		\mu \Phi _{1} &=&-\frac{1}{2}\left(\nabla _{r}^{2} -\frac{S_1^2}{r^2}\right) \Phi _{1}+\lambda \left( \frac{d}{dr}+\frac{S_2}{r}\right) \Phi
		_{2}-\Psi \Phi _{2},  \notag \\
		\mu \Phi _{2} &=&-\frac{1}{2}\left( \nabla _{r}^{2}-\frac{S_2^2}{r^{2}}\right) \Phi _{2}-\lambda\left(\frac{d}{dr}-\frac{S_1}{r} \right) \Phi _{1}-\Psi \Phi
		_{1},  \notag \\
		2\mu \Psi  &=&-\frac{1}{4}\left( \nabla _{r}^{2}
		-\frac{S_3^2}{r^{2}}\right) \Psi -\alpha \Psi - \Phi _{1}\Phi _{2}.  \label{PhiPsi}
	\end{eqnarray}%
	where $\nabla _{r}^{2}\equiv \partial _{r}^{2}+\frac{1}{r%
	}\partial _{r}$. At $r\rightarrow 0$, $\Phi _{2}$ and $\Psi $ vanish $\sim r$, while $\Phi
	_{1}$ remains finite. At $r\rightarrow \infty $, the asymptotic form of the
	SV soliton is a product of the exponentially decaying factor and oscillatory
	factors \cite{Sakaguchi}, $r^{-1/2}\exp \left( -\sqrt{-2\mu -\lambda ^{2}}r%
	\right) \cos \left( \lambda r+\delta \right) $, with a phase constant $%
	\delta $.

	We refer to the SV branch with $S_1=0$ as the GS family and the branches with $S_1\geq 1$ as ES families. Following Refs. \cite{HuangC,ZhongR}, the GS and ES solitons of SV type can be
	produced numerically by means of the imaginary-time integration method \cite%
	{Bao}, applied to Eqs. (\ref{+})-(\ref{++}) with the following input,
	written in the polar coordinates $\left( r,\theta \right) $:
	\begin{gather}
		\phi _{1}(t=0)=A_{1}r^{S_1}\exp (-\alpha _{1}r^{2}+iS_1\theta),  \notag \\
		\phi _{2}(t=0)=A_{2}r^{S_2}\exp (-\alpha _{2}r^{2}+iS_2\theta ),  \notag \\
		\psi (t=0)=\phi _{1}\phi _{2},  \label{input1}
	\end{gather}%
	where $A_{1,2}$ and $\alpha _{1,2}$ are positive real constants, and the
	molecular component of the ansatz is selected according to the
	phase-matching assumption.

	\section{Ground-state (GS) solitons}
	
\begin{figure}[tph]
		\center
		\includegraphics[height=1.7in,width=3.4in]{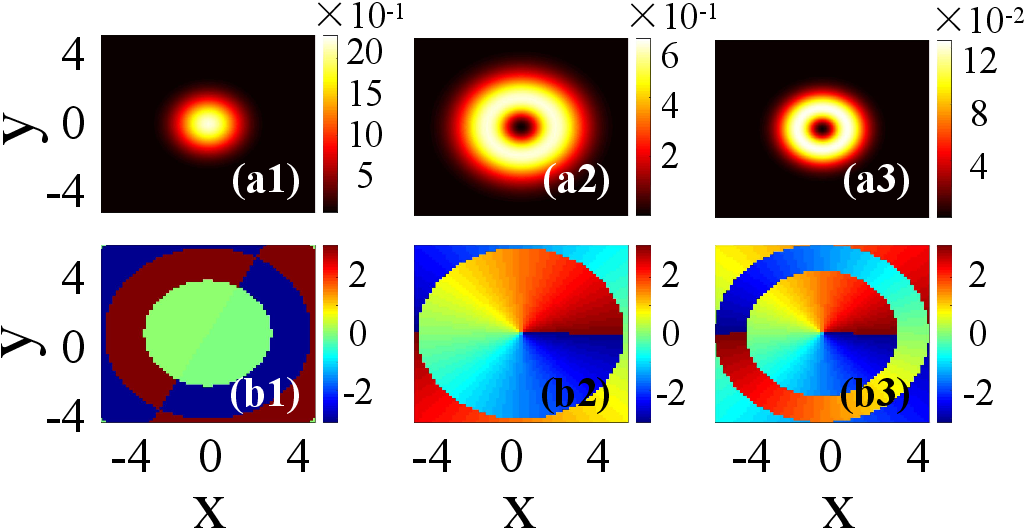}
		\caption{An example of a stable GS soliton, corresponding to
			ansatz (9) with $S_1= 0$. Panels (a1)-(a3) display, severally, the density patterns of the atomic
			components $\left\vert \protect\phi _{1}\right\vert ^{2}$ and $\left\vert
			\protect\phi _{2}\right\vert ^{2}$, and the molecular one, $\left\vert
			\protect\psi \right\vert ^{2}$. Panels (b1)-(b3) display, accordingly, phase
			patterns of the same components. The other parameters are $(N,\protect\lambda ,\protect\alpha )=(20,1,0)$.}
		\label{fig1}
	\end{figure}

	A typical example of a stable 2D GS soliton, with norm $N=20$ and
	parameters $\lambda =1$, $\alpha =0$ in Eqs. (\ref{+})-(\ref{++}) (no
	mismatch between the atomic and molecular components), is displayed in Fig. \ref{fig1}, where the density patterns of the
	atomic and molecular components ($\left\vert \phi _{1,2}\right\vert ^{2}$
	and $\left\vert \psi\right\vert ^{2}$, respectively), demonstrate
	perfect axisymmetric patterns, with vorticities $0$ in $\phi _{1}$ and $1$
	in $\phi _{2}$ and $\psi ,$ in agreement with the input given by Eq. (\ref%
	{input1}).
	
	\begin{figure}[tph]
		\center
		\includegraphics[height=2.1in,width=3in]{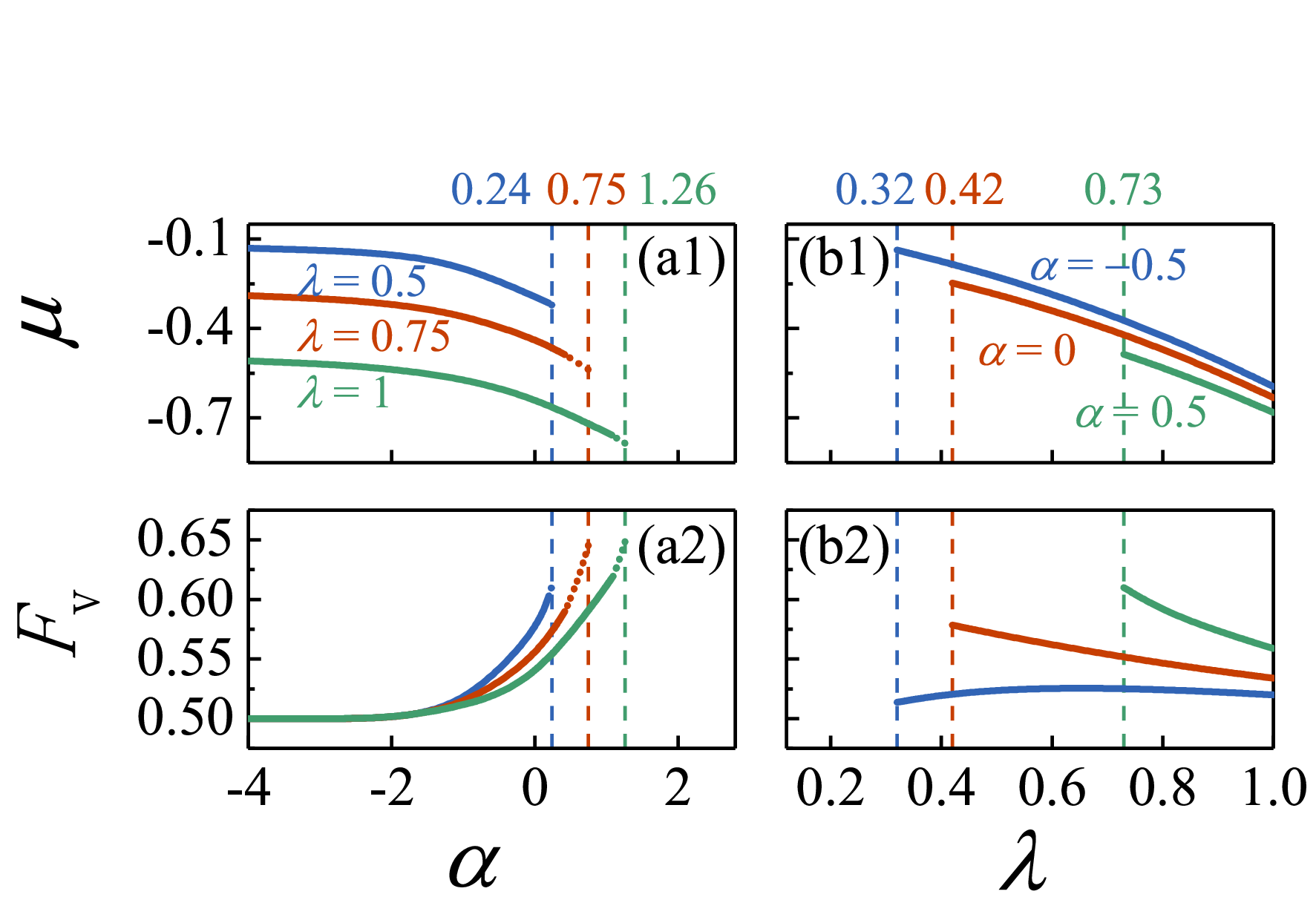}
		\caption{Panels (a1) and (a2) show, respectively, the chemical potential of
			the atomic components and the norm ratio of the vortex components $F_{v}$, vs. the detuning strength $\protect%
			\alpha $ for fixed values of the SOC $\protect\lambda =0.5$, $0.75$ and $1$%
			(blue, red, and green curves, respectively). The
			solid line corresponds to the parameter regime where solitons remain stable,
			while the dashed line demarcates the unstable region. The SV solitons exist
			with the detuning strength taking values below the threshold $\protect\alpha %
			_{c}$. Panels (b1) and (b2) show, respectively, the $\protect\mu $ and $F_{v}
			$ as functions of the SOC $\protect\lambda $ for $\protect\alpha =-0.5$, $0$
			and $0.5$(blue, red, and green curves,
			respectively). The SV solitons exist with the SOC strength taking values
			above the minimum threshold $\protect\lambda _{min}$.}
		\label{fig2}
	\end{figure}

	For GS soliton, vortex components are $\phi _{2}$
	and $\psi $, therefore the
	norm share of the vorticity-carrying components is defined as $F_{v}=\frac{N_{2}+N_{3}}{N}$. For a fixed value of the total norm, $N=20$, properties of the SV solitons
	are summarized in Figs. \ref{fig2}(a1,a2), where the chemical potential of
	atomic components and the norm ratio of the vortex components $F_{v}$, are
	plotted vs. the detuning strength $\alpha $ in Eqs. (\ref{+}), (\ref{-}) and
	(\ref{++}). With $\alpha <0$, the soliton's spatial extent grows
	significantly with the increase of $|\alpha |$, asymptotically approaching a
	spatially uniform solution, for which the chemical potential of atomic
	components attains its limiting value $\mu _{\max }=-\lambda ^{2}/2$.

	For different SOC strengths, there exists a critical value of $\alpha _{c}$, above which the SV solitons do not exist. This can be understood within the cascading-limit approximation, in which the rapidly varying molecular field is adiabatically eliminated \cite{C. Kong2026}, yielding an effective coupling term $\propto |\phi_j|^2/\alpha$ in the atomic equations: for $\alpha<0$ it provides an effective attraction that supports soliton formation, while for $\alpha>0$ it becomes repulsive and, above a threshold $\alpha_c$, overwhelms the SOC-induced confinement, destroying the SV soliton.

	For $N=20$, we also display the dependence of $\mu$ and $F_v$ as functions
	of the SOC strength $\lambda$ in Fig.\ref{fig2}(b1.b2). When $%
	\lambda<\lambda_{min}$, the SV soliton solutions do not exist, and the system
	tends to form MM solitons. When $\lambda>\lambda_{min}$, the SV
	soliton solutions exist, and the variation of the vortex population fraction
	with $\lambda$ reflects the stability of the solitons. In Fig. \ref{fig2}(b1), the values of $\mu$ gradually
	decrease, as the SOC strength increases. In Figs.\ref{fig2}(a2) and (b2),
	the norm ratio of the vortex components $F_v$ exceeds $0.5$.
	
\begin{figure}[tph]
		\center
		\includegraphics[height=1in,width=1.5in]{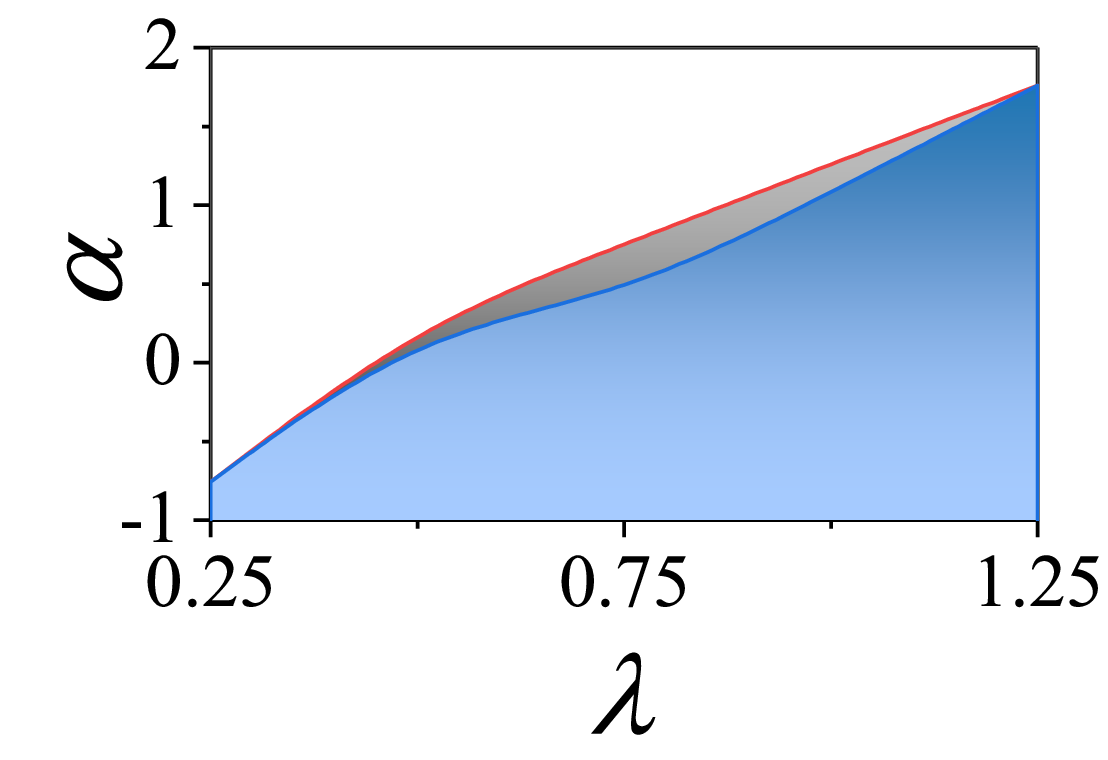}
		\caption{Parameter regions of the GS solitons in the $(\lambda, \alpha)$ plane are shown, with stable (blue), unstable (gray), and nonexistent (white) areas.}
		\label{fig3}
	\end{figure}

\begin{figure}[tph]
		\center
		\includegraphics[height=1.8in,width=3.4in]{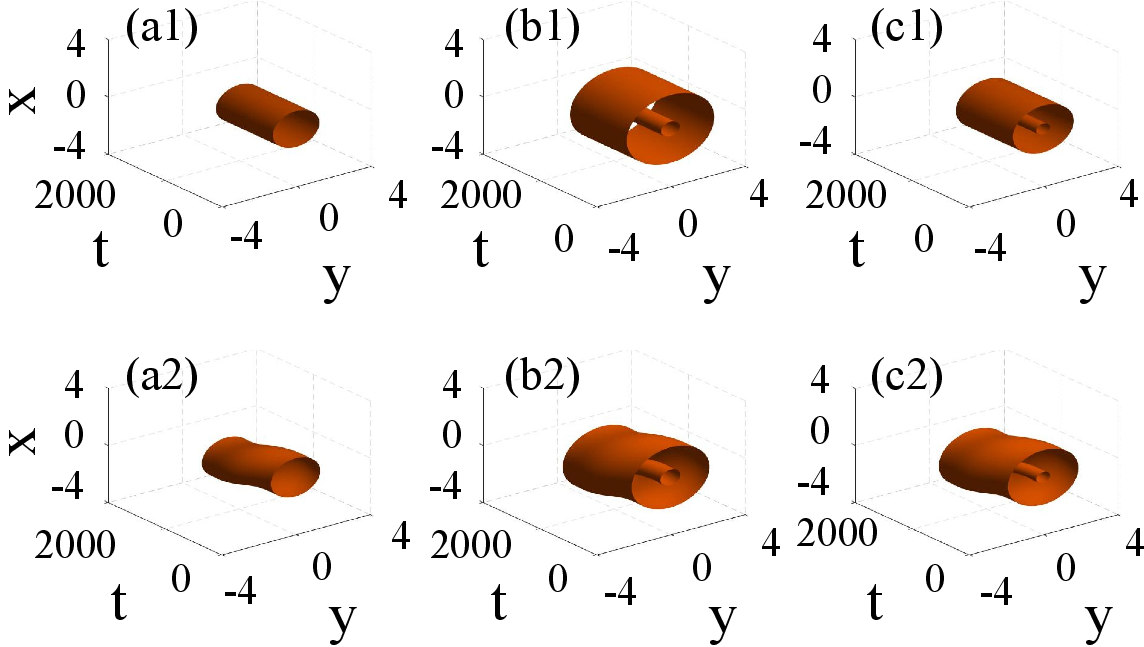}
		\caption{ Real-time evolution of GS soliton density profiles under $1\%$ random perturbations for $\lambda=1$,$N=20$. Panels (a1)–(c1) ($\alpha=0$, blue region in Fig. 3) show intact profiles over long-time evolution, confirming stability. Panels (a2)–(c2) ($\alpha=1.26$, gray region) show distortion, indicating instability.}
		\label{fig4}
	\end{figure}

	Parameter region of the GS solitons in the
	$(\lambda, \alpha)$ plane is plotted in Fig. \ref{fig3}. The
	SV solitons exist and are stable or unstable in the blue
	and gray areas, respectively, and they do not exist in the white region. The SOC $%
	\lambda $ and the detuning strength $\alpha$ play crucial regulatory roles
	in the atomic-molecular BEC system. To verify the stability of the GS solitons, we performed real-time simulations with $1\%$ random perturbations for two parameter sets corresponding to the stable (blue) and unstable (gray) regions in Fig. \ref{fig3}. With $\lambda = 1$, we set $\alpha = 0$ (blue region) and $\alpha = 1.26$ (gray region). As shown in Figs. \ref{fig4}(a1-c1), the density profiles for $\alpha = 0$ remain intact over thousands of time units, confirming robust stability. In contrast, for $\alpha = 1.26$ [Figs. \ref{fig4}(a2-c2)], the profiles suffer distortion, indicating instability. These results are fully consistent with the parameter regions of Fig. \ref{fig3}.
	
	The SOC breaks the Galilean invariance of GPEs,hence mobility of
	SOC-affected solitons is a nontrivial problem. Here, we define the
	transformed wave function and moving coordinates $\tilde{\phi}_{1,2 }= \exp
	\left( -\frac{i}{2}\left( V_{x}^{2}+V_{y}^{2}\right)
	t+iV_{x}x+iV_{y}y\right) \phi _{1,2}(\tilde{x},\tilde{y},t)$, $\tilde{\psi}=
	\exp \left( -i\left( V_{x}^{2}+V_{y}^{2}\right)
	t+2i\left(V_{x}x+V_{y}y\right)\right) \psi(\tilde{x},\tilde{y},t)$, $\tilde{%
		x}= x-V_{x}t$, and $y= \tilde{y}-V_{y}t$. After applying the Galilean
	transformation, an additional term, $(iV_{x}+V_{y})\tilde{\phi}_{2}$ and $%
	(-iV_{x}+V_{y})\tilde{\phi}_{1}$, is introduced to Eqs. (\ref{+}) and (\ref%
	{-}) respectively. Eq. (\ref{++}) remains unchanged in form due to its
	compliance with Galilean invariance. The numerical results indicate that the
	system under a moving reference frame only supports the MM solutions, and
	the SV solutions do not exist.

	\section{Excited-state(ES)solitons}
	
	For an ES soliton generated by input Eq. (7) with a given integer $S_1 \ge 1$, the three components carry distinct topological charges that strictly satisfy the phase-matching condition imposed by the quadratic nonlinearity. Consequently, unlike the GS SV, where only $\phi_2$ and $\psi$ are vortices while $\phi_1$ is a fundamental mode, every component of the ES soliton carries a non-zero topological charge. In the present 2D system, the ES solitons (with topological charges $S_1 \ge 1$) are genuine pure vortices, meaning that the entire self-bound state carries a single dominant topological charge without the admixture of a fundamental-mode component. This qualitative change, which is absent in the 3D counterpart, originates from the more favorable geometric constraints in 2D that allow the SOC to fully lock the vorticity across both atomic and molecular components. This pure-vortex characteristic constitutes a key innovation of the 2D configuration relative to our previous 3D work.

\begin{figure}[tph]
		\center
		\includegraphics[height=4in,width=3.4in]{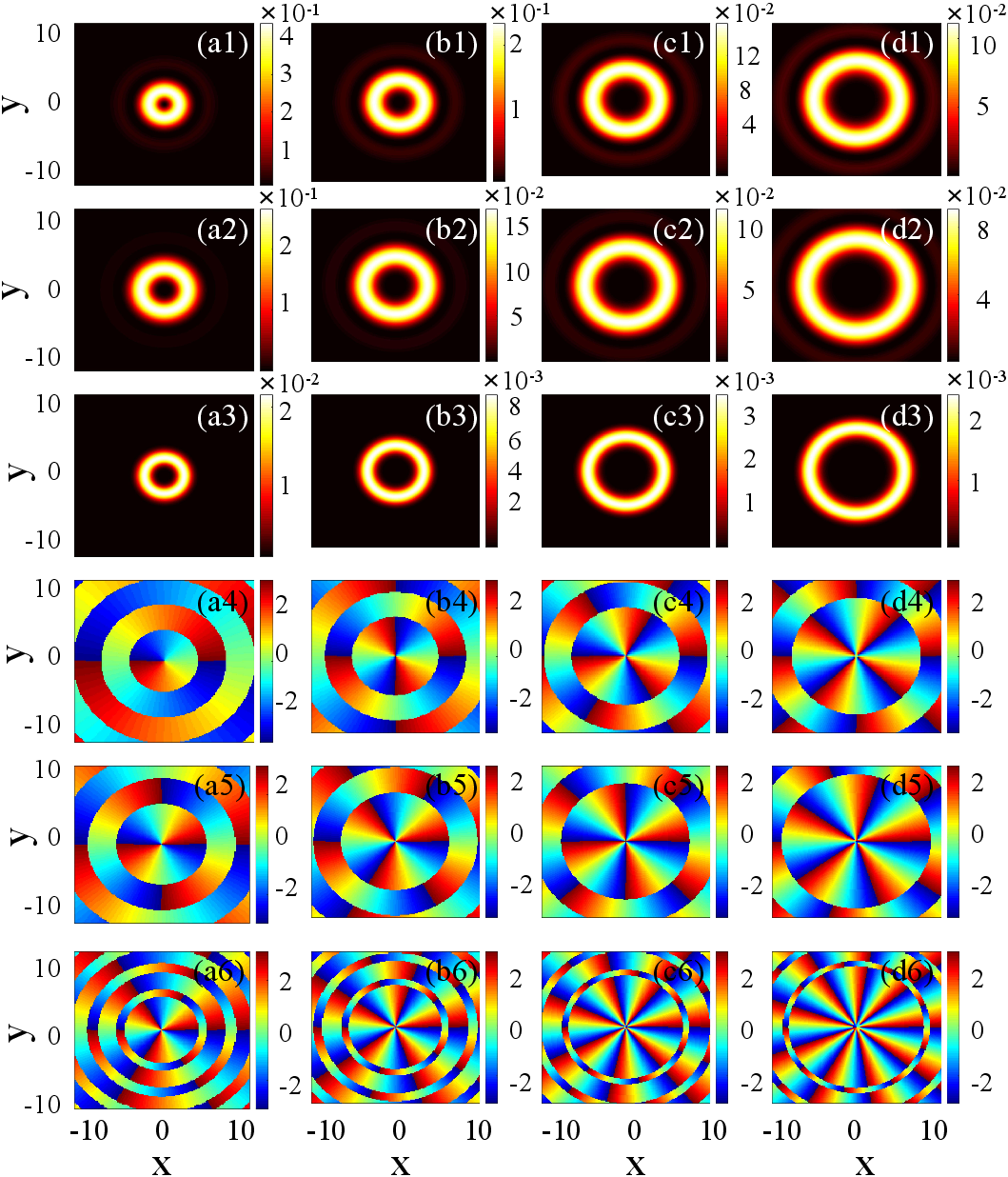} %
		\caption{Panels display the typical examples of stable 2D ES solitons of SV type, with parameters $(N,\protect\lambda ,\protect\alpha )=(20,1,0)$.
			Panels (a1-d1), (a2-d2) and (a3-d3) show, respectively, the density patterns of the atomic
			components $\left\vert \protect\phi _{1}\right\vert ^{2}$, $\left\vert
			\protect\phi _{2}\right\vert ^{2}$, and the molecular one, $\left\vert
			\protect\psi \right\vert ^{2}$. Panels (a4-d4), (a5-d5) and (a6-d6) display, accordingly, phase
			patterns of the same components. From left
			to right: $S_1 = 1, 2, 3$, and $4$.}
		\label{fig5}
	\end{figure}

	Using the imaginary-time evolution method with input Eq. (7), we have constructed stable ES soliton solutions for topological charges up to $S_1=5$. Typical examples of the density and phase structures for $S_1=1,2,3,$ and $4$ are displayed in Fig. \ref{fig5} (with $N=20$, $\lambda=1$, and $\alpha=0$). These profiles clearly show the concentric ring patterns characteristic of high-order vortices, with the phase winding increasing monotonically with $S_1$.
	
	\begin{figure}[tph]
		\center
		\includegraphics[height=2.6in,width=3.2in]{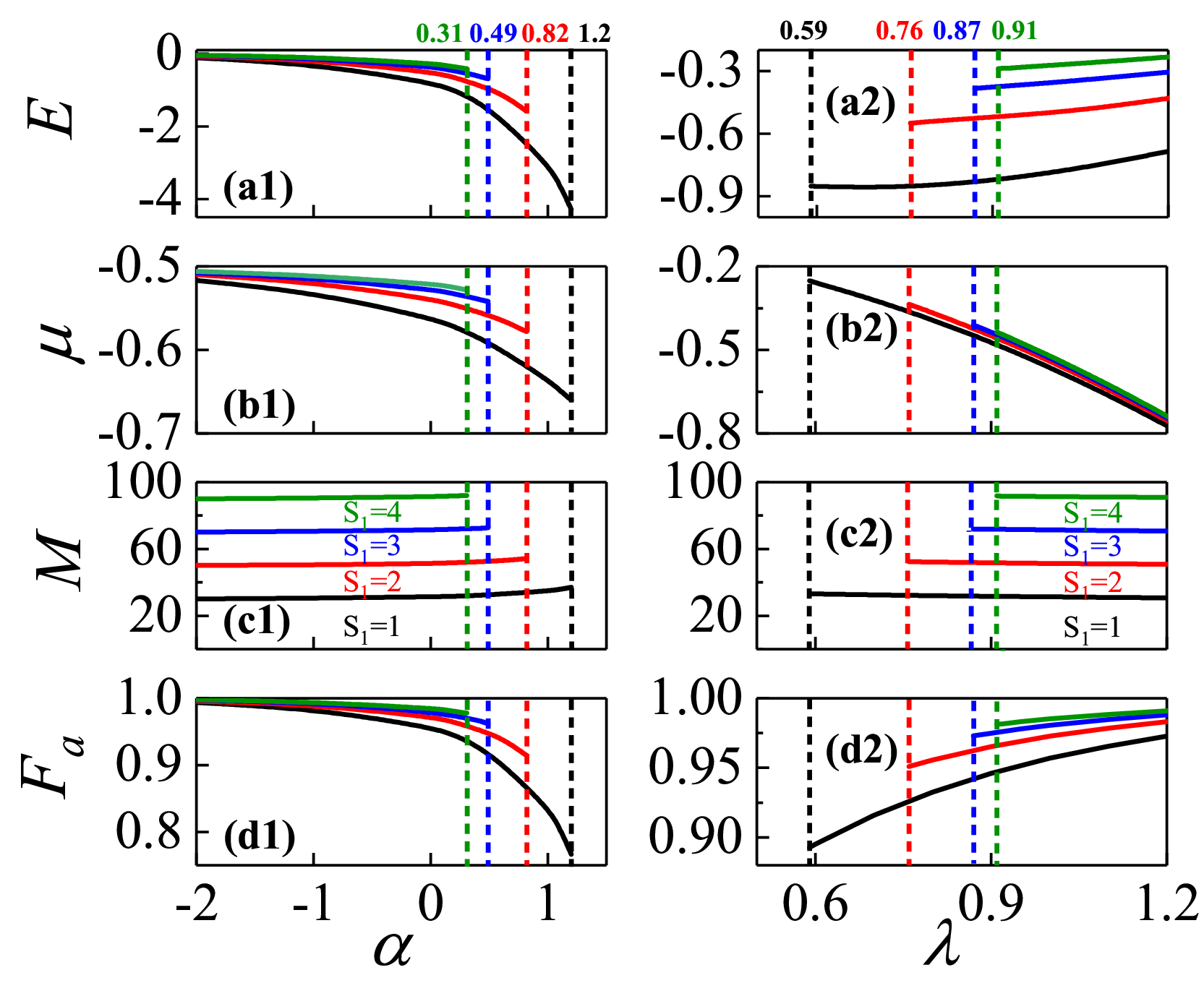} %
		\caption{The dependence of the ES-SV soliton energy $E$, chemical potential $\mu$, angular momentum $M$ and atomic fraction $F_a = (N_1+N_2)/N$ on detuning $\alpha$ [panels (a1-d1), with $\lambda = 1$], and on SOC strength $\lambda$ [panels (a2-d2), with $\alpha=0$]. From bottom to top are $S_1 = 1, 2, 3$, and $4$.}
		\label{fig6}
	\end{figure}

\begin{figure}[tph]
		\center
		\includegraphics[height=1.09in,width=3.2in]{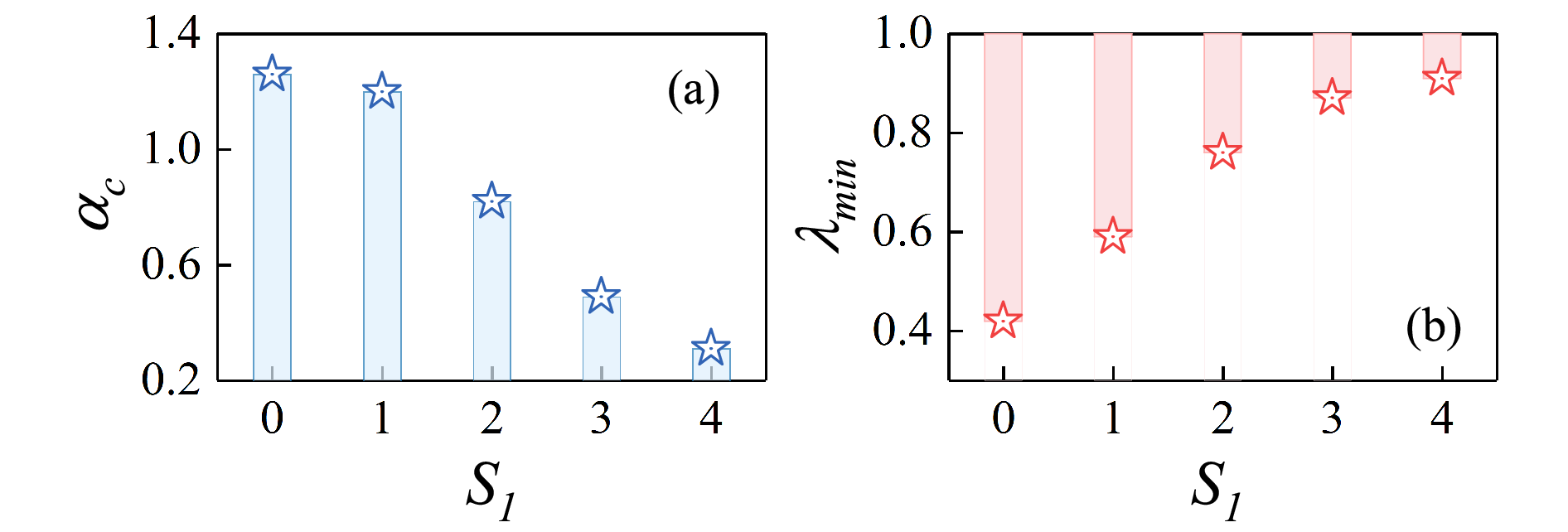} %
		\caption{ (a) Critical detuning $\alpha_c$ versus $S_1$ with $\lambda = 1$, showing a systematic decrease with $S_1$. SV soliton solutions exist when $\alpha<\alpha_c$. (b) Minimum SOC strength required for stabilizing the SV solitons versus $S_1$ with $\alpha = 0$, exhibiting a monotonic increase. SV soliton solutions exist when $\lambda>\lambda_{\min}$.}
		\label{fig7}
	\end{figure}

	Figure \ref{fig6} presents a comprehensive characterization of the ES-SV solitons through their chemical potential $\mu$, energy $E$, and angular momentum $M$ as functions of the detuning $\alpha$ [panels (a1)–(c1), with $\lambda=1$] and of the SOC strength $\lambda$ [panels (a2)–(c2), with $\alpha=0$]. As shown in panels (a1) and (b1), for each topological charge $S_1$, both $\mu$ and $E$ decrease monotonically with increasing $\alpha$. These trends characterize the parametric evolution of the stationary ES solutions. Panel (c1) further reveals that $M$ increases monotonically with $\alpha$ for each $S_1$. To understand the increase of $M$ with $\alpha$, we analyze the internal redistribution of the total norm among the three components. We define the atomic fraction $F_a = (N_1 + N_2)/N$ and the molecular fraction $F_m = N_3/N = 1 - F_a$. Fig. \ref{fig6}(d1) shows that $F_a$ decreases monotonically with increasing $\alpha$ (so that $F_m$ increases correspondingly). Since the molecular field carries the highest topological charge $S_3 = 2S_1 + 1$ while the atomic components carry only $S_1$ and $S_2 = S_1 + 1$, this population transfer from the atomic to the molecular component enhances the overall angular momentum $M$, consistently explaining the monotonic increase of $M$ with $\alpha$. We now turn to the quantitative thresholds that bound the stability regime, as shown in Figure \ref{fig7}(a). For the GS soliton (corresponding to $S_1=0$), the critical detuning is $\alpha_c = 1.26$, which marks the existence boundary: for $\alpha > 1.26$, the GS solution ceases to exist; for $\alpha < 1.26$, the GS solution exists, but its stability is not uniform (there exist subregions where it is stable and others where it is unstable, as detailed in Sec. III). For the ES solitons ($S_1=1,2,3,4$), stability is maintained only for detuning values below their respective critical values $\alpha_c$, which decrease systematically with $S_1$: we find $\alpha_c = 1.2, 0.82, 0.49$, and $0.31$ for $S_1 = 1, 2, 3$, and $4$, respectively. Beyond these thresholds, the ES solitons cease to exist and transition into MM states. These critical detunings can be understood within the cascading-limit approximation: as $\alpha$ increases toward $\alpha_c$, the effective nonlinearity weakens, eventually losing the balance required for self-trapping. This monotonic decrease of $\alpha_c$
	with increasing topological charge further confirms that higher-order vortices are more susceptible to detuning-induced destabilization.

	Fig. \ref{fig6}(a2-c2) illustrates the analogous dependencies on $\lambda$, with parameters fixed at $N=20$ and $\alpha=0$. As shown in panel (a2), for each topological charge $S_1$, the energy $E$ increases monotonically with increasing $\lambda$, satisfying $dE/d\lambda>0$ throughout the entire parameter regime, indicating that stronger SOC raises the system energy. In contrast, panel (b2) shows that the chemical potential $\mu$ decreases monotonically as $\lambda$ increases. In panel (c2), the total angular momentum $M$ also decreases monotonically with increasing $\lambda$, suggesting that although SOC enhances the vorticity-locking effect, it simultaneously redistributes the population among the three components in a manner that reduces the overall angular momentum. To understand this decreasing trend, we  display the evolution of $F_a$ as a function of $\lambda$ for different topological charges $S_1$. As shown in panel (d2), $F_a$ increases monotonically with $\lambda$, implying that $F_m$ decreases correspondingly. Since the molecular component carries the highest topological charge, the molecular field dominates the contribution to the total angular momentum. Consequently, the reduction of the molecular population inevitably lowers the overall angular momentum $M$, despite the enhanced SOC-induced vorticity locking. This population transfer mechanism thus provides a direct microscopic interpretation for the observed $dM/d\lambda < 0$. 
	
	A crucial finding concerns the existence threshold: for each $S_1$, there exists a minimum SOC strength $\lambda_{min}$
	: when $\lambda < \lambda_{min}$, the ES-SV soliton solutions cease to exist and the system tends to form MM solitons; when $\lambda > \lambda_{min}$, the ES-SV soliton solutions emerge and become progressively more robust with increasing $\lambda$. As systematically determined in Fig. \ref{fig7}(b), $\lambda_{\min}$ increases monotonically with the topological charge: for $S_1 = 0, 1, 2, 3$, and $4$, we find $\lambda_{\min} = 0.42, 0.59, 0.76, 0.87$, and $0.91$, respectively. This monotonic rise reflects that higher-order vortices carry larger angular momentum and therefore require stronger SOC to overcome the azimuthal instability and maintain rigid phase coherence across all components. Notably, this threshold behavior highlights the pivotal role of SOC in enabling fully quantized vortex configurations, thereby resolving the incompleteness encountered in previous 3D studies where the vortex fraction remained below $100\%$.
	
	To connect with realistic experimental conditions, we consider a
	$^{39}$K condensate with scattering length $a_s\simeq 5a_0$, peak 3D density
	$n_{3D}\simeq 10^{20}\,\mathrm{m^{-3}}$, and transverse trapping frequency
	$\omega_z\simeq 2\pi\times 1\,\mathrm{kHz}$ \cite{Roati2007,CDM2021,Sadler2006}. The corresponding 2D areal
	density is $n_{2D}=n_{3D}\sqrt{\pi}a_z=9.0\times10^{13}\,\mathrm{m^{-2}}$,
	with $a_z=\sqrt{\hbar/(m\omega_z)}=0.51\,\mu\mathrm{m}$. These parameters
	yield the characteristic length $L_0=1.22\,\mu\mathrm{m}$, energy
	$E_0/h\simeq87\,\mathrm{Hz}$, and time $t_0=\hbar/E_0=1.82\,\mathrm{ms}$.
	The scaled particle number $N$ relates to the physical atom number by
	$N_{\mathrm{phys}}=Nn_{2D}L_0^2$, so $N=20$ corresponds to
	$\approx 2.7\times10^3$ atoms. The dimensionless detuning $\alpha=1$ and
	SOC strength $\lambda=1$ correspond to physical frequencies
	$\nu_\alpha=\nu_\lambda=E_0/h\simeq87\,\mathrm{Hz}$. The dimensionless
	evolution time $t=1000$ corresponds to $t_{\mathrm{phys}}\approx 1.82\,
	\mathrm{s}$, which is sufficiently long for observing the self-bound states.
	These estimates suggest that the characteristic density, atom number, length, and time
	scales considered here are compatible with current cold-atom experiments.

	\section{Conclusion}
	
	We systematically investigate the possibility of achieving stable vortex solitons in a 2D atomic-molecular BEC with quadratic nonlinearity and detuning, by introducing SOC exclusively to the atomic component. A central motivation is to overcome the limitation encountered in previous 3D studies, where the vortex fraction remained incomplete and a fully quantized vortex configuration could not be realized. By reducing the system to two dimensions, we demonstrate that fully self-bound vortex states become accessible. A crucial finding concerns the existence threshold of these soliton states. For both GS and ES solitons, there exists a minimum SOC strength required for their emergence, which increases monotonically with the topological charge. Below this threshold, the system fails to support vortex solitons and instead forms MM states; above it, the soliton solutions emerge and become progressively more robust with increasing SOC. This systematic elevation of the threshold reflects that higher-order vortices carry larger angular momentum and consequently require stronger SOC to overcome the azimuthal instability. The critical detuning, similarly, decreases monotonically with the topological charge, indicating that higher-order states are more susceptible to detuning-induced destabilization. These findings provide a new theoretical pathway for generating and stabilizing high-vorticity solitons in quadratic nonlinear systems, and highlight the role of SOC in enabling fully quantized vortex configurations that were previously unattainable in 3D settings.
	
	A natural direction for further research is to extend the
	current analysis for ES solitons of the MM
	type, which combine terms with vorticities $0$ and $\pm1$ in the
	two components. In particular, the asymmetric vortex structures and the competition between different internal modes may open up new stability regimes, and clarifying these mechanisms would deepen our understanding of complex self‑trapped states in 2D SO‑coupled BECs.

	\emph{Acknowledgments-}This work was supported by the Project of the National Natural Science
	Foundation of China under Grants No. 12505031 and No. 12675027, the Project of Hunan Provincial Education Office under Grant No. 25C0624, Guangdong Basic and Applied
	Basic Research Foundation under Grants No. 2024A1515010710 and No. 2026A1515012340, Scientific Research
	Foundation of Xiangnan University for High-Level Talents, the Applied
	Characteristic Disciplines of Electronic Science and Technology of Xiangnan
	University, and Hunan Provincial Innovation Training Program for College Students (Document No. [2026]108).

\end{document}